\documentstyle[12pt,epsf]{article}
\begin{document}
\baselineskip=24pt
\pagestyle{plain}

\begin{center}
{\Large \bf Classical and Quantum Solutions of 2+1 Dimensional Gravity
            under De Broglie-Bohm Interpretation}
\\
\vspace{1cm}
{\large M. Kenmoku
\footnote[1]{
kenmoku@phys.nara-wu.ac.jp},
T. Matsuyama \footnote[2]{matsuyat@nara-edu.ac.jp},
R. Sato \footnote[3]{reika@phys.nara-wu.ac.jp}
and S. Uchida \footnote[4]{satoko@phys.nara-wu.ac.jp}}
\\
\vspace{0.5cm}
{\small \it \footnotemark[1]
Department of Physics, Nara Women's University,
Nara 630-8506, Japan \\
\footnotemark[2]
Department of Physics, Nara University of Education, Takabatake-cho,
Nara 630-8528, Japan \\
\footnotemark[3] \footnotemark[4]
Graduate School of Human Culture, Nara Women's University,
Nara 630-8506, Japan}
\end{center}

\vspace{2cm}

\begin{abstract}
We have studied classical and quantum solutions of 2+1 dimensional
Einstein gravity theory.
Quantum theory is defined through the local conserved angular momentum
and mass operators in the case of spherically symmetric space-time.
The de Broglie-Bohm interpretation is applied for the wave function
and we derive the differential equations for metrics.
Metrics including quantum effect are obtained in solving these equations
and we compare them with classical metrics.
Especially the quantum effect on the closed de Sitter universe is 
evaluated quantitatively. 
\end{abstract}

\newpage

\section{Introduction}


Recently 2+1 dimensional gravity theory, especially ${\rm AdS_3}$
has been studied extensively \cite{DJT84,BTZ92,Kubota99}.
It was shown that the theory is equivalent to
the 2+1 Chern-Simons theory \cite{Witten88}
and has been investigated to understand
the black hole thermodynamics, i.e. Hawking temperature \cite{Hawking75} and others. 
The 2+1 dimensional AdS supergravity theories were 
investigated in the Chern-Simons form \cite{Achucarro86}.
Topalogically massive gravity, which is the sum of 
Einstein and Chern-Simons terms, was also 
analyzed in the canonical form \cite{Deser91}. 
The purpose of this paper is to investigate the canonical formalism of
the original 2+1 dimensional Einstein gravity theory
instead of the Chern-Simons theory.
For the spherically symmetric space-time,
local conserved quantities (mass and angular momentum) are
introduced and using them canonical quantum theory is defined.
Constraints are imposed on state vectors and solved analytically. 
In order to extract the physical meaning of the wave function,
we adopt the de Broglie-Bohm interpretation \cite{Bohm52,Bell87,Holland93}
and derive the differential equations for the metrics,
which include the quantum effect.


After fixing the gauge choice, special solutions of the metrics
are obtained.
Especially the quantum effect on the closed de Sitter universe is
obtained quantitatively.  It is interesting to note that
the birth of universe appears
as the real time tunneling in stead of the imaginary time tunneling
in the WKB approximation by Vilenkin \cite{Vilenkin82}
and in the path integral method by Hartle and Hawking \cite{HH83}.
The strategy to obtain the solution is followed by
our previous work in the 3+1 dimensional gravity \cite{KKTY99}. 
The new developments in this paper are 
to take into account a general form of metrics and 
to make a quantitative evaluation to quantum effects 
in the 2+1 dimensional spherically symmetric gravity. 


This paper is organized as follows.
In section  2, we review the canonical formalism of
the Einstein theory with a cosmological constant
in spherically symmetric space-time. We introduce
the local conservation quantities,
the angular momentum \(J\) and the mass function \(M\).
In section 3, the canonical quantization is presented.
Using some operator ordering the quantum theory could be
constructed and the analytic solutions could be obtained.
In section 4, we adopt the de Broglie-Bohm interpretation
in order to extract geometrical information from
the analytic solutions. 
Some examples of the comparison between classical limit and
classical solutions which have already been known 
are also exhibited.
In section 5, 
we try to evaluate the finite quantum effect 
for closed de Sitter universe. 
Summary and concluding remark are given in the final section.

\section{Canonical Formalism}

We start to consider the Einstein-Hilbert action with cosmological 
constant
\( \lambda \)  in 2+1 dimensional space-time,
\begin{equation}
I={1 \over 16 \pi G_2} \int d^{3}x \sqrt{-^{(3)}g} (^{(3)}R-2\lambda) .
\label{action}
\end{equation}
The gravitational constant in 2+1 dimensions is set to
$G_2=1/4$ in the following.
The metrics in polar coordinates are expressed in ADM
decomposition \cite{ADM62} as
\begin{eqnarray}
ds^{2}=-N^{2}dt^{2}+\Lambda^{2}(dr+N^{r}dt)^{2}
       +R^{2}(d\phi+N^{\phi}dt)^{2}
\\ \nonumber
+2C(dr+N^{r}dt)(d\phi+N^{\phi}dt) ,
\label{metric}
\end{eqnarray}
where all metrics are assumed to be function of time $t$ and radial
coordinate $r$. In the following, dot and dash denotes the derivative
with respect to $t$ and $r$.

The action in the canonical formalism is in the form,
\begin{eqnarray}
I &=&  \int dt \, dr \;[P_{\Lambda}\dot{\Lambda}+P_{R}\dot{R}
                                                +P_{C}\dot{C}
-(NH + N^{r} H_{r} + N^{\phi} H_{\phi})]
\nonumber \\
& & -  \int dt \, dr \biggl( [(\Lambda P_{\Lambda} 
    + C P_{C})N^{r} ]^{\prime}
+  [ ({C \over \Lambda} P_{\Lambda} + R^{2} P_{C}) N^{\phi} ]^{\prime}
\biggl) \;,
\end{eqnarray}
where canonical momenta are
\begin{eqnarray}
P_{\Lambda} &=& {\partial {\cal L} \over \partial \dot{\Lambda}}=
{\Lambda R (N^{r} R^{\prime}-\dot{R}) \over N \sqrt{h}} \;,
\label{clmomLambda} \\
P_{R} &=& {\partial {\cal L} \over \partial \dot{R}}=
{R[C{N^{\phi}}^{\prime}+\Lambda \{(\Lambda N^{r})^{\prime}
- \dot{\Lambda} \}]
\over N \sqrt{h}} \;,
\label{clmomR} \\
P_{C} &=& {\partial {\cal L} \over \partial \dot{C}}=
-{(N^{r}C)^{\prime} + R^{2} {N^{\phi}}^{\prime} - \dot{C} \over
2 N \sqrt{h} } \;,
\label{clmomC}
\end{eqnarray}
and the Hamiltonian and the momentum functions are defined as
\begin{eqnarray}
H &=& \sqrt{h}  \biggl(- {P_{\Lambda}P_{R} \over \Lambda R} 
+ P_{C}^{2} + 
\frac{ {R^{\prime}}^{2}+ 
R R^{\prime \prime}}{h}
-{R R^{\prime} h^{\prime} \over 2h^2} + \lambda \biggl) \;,
\\
H_{r}&=&  R^{\prime}P_{R}-C {P_{C}}^{\prime}
-\Lambda {P_{\Lambda}}^{\prime}  \;,
\\
H_{\phi}&=& - \biggl({C \over \Lambda} P_{\Lambda}
+R^{2} P_{C} \biggl)^{\prime}  \;.
\end{eqnarray}
In the expression of canonical momenta 
in Eq. (\ref{clmomLambda})-(\ref{clmomC}), 
${\cal L}$ is the Lagrangian density and $h=\Lambda^2R^2-C^2$ is 
the determinant of the spatial part of metrics. 

It is essential to introduce the local conservation quantities,
the angular momentum \( J \) and the mass function \( M \)
as follows 
\begin{eqnarray}
J &:=& - 2 \int dr H_{\phi} = 
2 \biggl( {C \over \Lambda} P_{\Lambda}+R^{2} P_{C} \biggl) \;,
\label{angular} \\
M &:=&
- \int dr \biggl( {R R^{\prime} \over \sqrt{h} }H + {P_{\Lambda} \over
\Lambda}H_{r}
+ P_{C} H_{\phi} \biggl) \nonumber
\\ &=&
{1 \over 2} \biggl({P_{\Lambda}}^{2} + {2C P_{\Lambda} P_{C} \over \Lambda}
+ R^{2} {P_{C}}^{2} - { R^2 {R^{\prime}}^{2} \over h }
- \lambda R^{2}  +1 \biggl)\;.
\label{mass}
\end{eqnarray}
The angular momentum function 
\(J\) could be derived as a conservative quantity
for a infinitesimal continuous rotation by Noether's Theorem,
and the mass function \(M\) could be derived by using 
the form of \(J\) and 
the transformation method from the canonical data 
which was considered by 
Fischler, Morgan and Polchinski \cite{bib:wfdmjp90}
and by Kucha\u{r} \cite{bib:kuch90}.
The physical meaning of the mass function 
in the 3+1 dimensional space-time is studied by Nambu
and Sasaki \cite{bib:ynms88}. 
These local conserved quantities are essential in defining 
and solving the quantum theory in spherically symmetric gravity.

Before going to the detail of the quantum theory, 
we make transformation from old variables \( \Lambda, R\) and \( C \) into
new variables by
\begin{equation}
\left( \begin{array}{c} \Lambda \\ R \\ C \end{array} \right)
\longrightarrow
\left( \begin{array}{c} \ \bar{\Lambda} \\ \bar{R} \\ \bar{C} \end{array}
\right)
= \left( \begin{array}{c} \sqrt{\Lambda^{2}-{C^{2} R^{-2}}} \\ R
\\ {C  R^{-2}} \end{array} \right) .
\label{pointra}
\end{equation}
The corresponding momenta are transformed as
\begin{eqnarray}
\left( \begin{array}{c} P_{\Lambda} \\ P_{R} \\ P_{C} \end{array} \right)
\longrightarrow
\left( \begin{array}{c} P_{\bar{\Lambda}} \\ P_{\bar{R}} \\ P_{\bar{C}}
\end{array} \right)
= \left( \begin{array}{c} \sqrt{\Lambda^{2}-{C^{2} R^{-2}}} 
 \Lambda^{-1} P_{\Lambda} \\
{C^{2} \Lambda^{-1} R^{-3}} P_{\Lambda} + P_{R} + 2C R^{-1} P_{C}
\\ C \Lambda^{-1} P_{\Lambda} + R^{2} P_{C} \end{array} \right) .
\label{canmom}
\end{eqnarray}
These new variables will be used extensively 
in the following calculation. 

\section{Quantum Theory} \label{sec:quanthe}

Next we proceed to the quantum theory 
in the Schr${\rm \ddot{o}}$dinger picture. 
The momentum operators are expressed in the new variables 
in Eqs.(\ref{pointra}) and (\ref{canmom}) as 
\begin{eqnarray}
\hat{P}_{\bar \Lambda}(r) 
:= -i {\delta \over \delta \bar \Lambda(r)} \ , \ 
\hat{P}_{\bar R}(r) := -i {\delta \over \delta \bar R(r)} \ , \ 
\hat{P}_{\bar C}(r) := -i {\delta \over \delta \bar C(r)} \ . 
\end{eqnarray}
The notation hat denotes the quantized operator in the followings. 
According to the Dirac approach, the constraints are treated
as operator restriction on the wave function \(\Psi\): 
\begin{equation}
\hat{H} \Psi=0 \;\;,\;\;
\hat{H}_{r} \Psi =0 \;\;,\;\;
\hat{H}_{\phi} \Psi =0 \;\;.
\label{wd}
\end{equation}
The first equation is the Wheeler-DeWitt equation and others 
are the momentum constraint equations. 


Our strategy is to solve the eigenvalue equation
for $\hat{J}$, $\hat{M}$ and the constraint equation
for $\hat H_r$ step by step instead of solving the constraint equations 
(Eq. (\ref{wd})). 
\\
Step 1: Angular momentum eigenequation \\
The eigenvalue equation of the local angular momentum
(Eqs. (\ref{angular}) and (\ref{canmom})) with the eigenvalue $j$
\begin{eqnarray}
\hat{J} \Psi = \hat{P_{\bar{C}}} \Psi = j \Psi \;
\label{step1}
\end{eqnarray}
is satisfied by
the eigenfunction in the form
\begin{eqnarray}
\Psi = e^{ij \Phi} u(\bar{\Lambda},\bar{R}) \;,
\end{eqnarray}
with the variable 
\begin{eqnarray}
\Phi = \int dr \bar{C}(r) \;.
\end{eqnarray}
\noindent
Step 2: Momentum constraint equation \\
The radial momentum constraint equation
\begin{eqnarray}
\hat{H}_{r} \Psi =
 (\bar{R}^{\prime} \hat{P}_{\bar{R}} -
 \bar{\Lambda} (\hat{P}_{\bar{\Lambda}})^{\prime} )
 e^{ij\Phi} u(\bar{\Lambda},\bar{R})
= 0 \;,
\label{step2}
\end{eqnarray}
restricts the functional form of the wave function to
\begin{equation}
\Psi = e^{ij \Phi} u(Z) \;,
\end{equation}
where we introduce variable \( Z \)
\begin{eqnarray}
Z = \int dr \bar{\Lambda} f(\bar{R},\chi)
  = \int dr \int^{\bar{\Lambda}(r)} d\bar{\Lambda}
                  \bar{f} (\bar{R},\chi) \;,
\label{zett}
\end{eqnarray}
with the definition 
\begin{eqnarray}
\chi := {{\bar{R}\,^{\prime}}\,^{2} \bar{\Lambda}^{-2}} \;.
\label{chi}
\end{eqnarray}
The function $f$ and $\bar{f}$ are arbitrary functions and 
are related each other by
\begin{eqnarray}
f(\bar{R},\chi) &=& -\int^{\chi} d\chi {\bar{f}(\bar{R},\chi) \over 2\chi}
\;.
\end{eqnarray}
\nonumber
Step 3:  Mass equation \\ 
The mass operator $\hat M$ consists of the integration 
of the linear combination of constraints 
with respect to radial coordinate $r$ in Eq.(\ref{mass}) 
and is imposed on the state vector as
\begin{eqnarray}
(\hat M -m)\Psi =0 \;, 
\label{step3}
\end{eqnarray}
where m is an integration constant. 
Three operators $\hat J$ , $\hat{M}$ and $\hat{H}_r$ 
should form a closed algebra in order to ensure 
the consistency among three wave equations 
(\ref{step1}), (\ref{step2}) and (\ref{step3}): 
\begin{eqnarray}
\, [\hat{J}(r),\hat{J}(r^{\prime})] &=& 0 \;,
\\
\, [ \hat{J}(r),{\hat{H}}_{r}(r^{\prime})]
&=& i \hat{J}^{\prime}(r) \delta(r-r^{\prime}) \;,
\\
\, [\hat{J}(r),\hat{M}(r^{\prime})] &=& 0 \;,
\\
\, [{\hat{H}}_{r}(r),{\hat{H}}_{r}(r^{\prime})]
&=& i ({\hat{H}}_{r}(r) \delta^{\prime}(r-r^{\prime})
-(r \leftrightarrow r^{\prime}) ) \;,
\\
\, [{\hat{H}}_{r}(r),\hat{M}(r^{\prime})]
&=& i \hat{M}^{\prime}(r) \delta(r-r^{\prime}) \;,
\\
\, [\hat{M}(r),\hat{M}(r^{\prime})] &=& 0 \;. 
\end{eqnarray}
The factor ordering of mass operator \( \hat{M} \) 
is determined by this requirement and is expressed as 
\begin{equation}
\hat{M} -m = {1 \over 2}
            A \hat{P}_{\bar{\Lambda}}A^{-1} \hat{P}_{\bar{\Lambda}}
        + {1 \over 2} ( -\chi + \hat{F}(\bar{R})) \;,
\end{equation}
where $\chi$ is in Eq.(\ref{chi}) and 
\begin{eqnarray}
\hat{F}(\bar{R}) =  1-2m-\lambda \bar{R}^{2} +
                    {1 \over 4} {\hat{J}^{2} \bar{R}^{-2}} \;.
\end{eqnarray}
The ordering factor $A$ is introduced to take account of 
the factor ordering ambiguity and is expressed by the product 
of two functions as 
\begin{equation}
A = A_{Z}(Z) \bar{A}(\bar{R},\chi) \; ,  
\end{equation}
where $A_{Z}$ and $\bar{A}$ are arbitrary functions with respect to 
the variable $Z$ and $\bar{R}\;, \;\chi$ respectively. 
We choose one of these function $\bar{A}$ as 
\begin{equation}
\bar{A} = {\delta Z \over \delta \bar{\Lambda}} = \bar{f}
        = \sqrt{\chi - F_j(\bar{R})} \;,
\label{fbar}
\end{equation}
where
\begin{equation}
F_j(\bar R):= \hat{F}\mid_{\hat{J}=j} \label{fj} (\bar R)\ .
\end{equation}
Then using the mass operator for each eigenvalue of angular momentum $j$
\begin{equation}
\hat{M}_{j}:= \hat{M}\mid_{\hat{J}=j} \ ,
\end{equation}
the mass equation
\begin{equation}
\hat{M}_{j} u_{j,m}(Z)= m u_{j,m}(Z) \; ,
\end{equation}
can reduce to the equation with respect to $Z$
\begin{equation}
{d^{2}u_{j,m}(Z) \over dZ^{2}}
- {A_{Z}}^{-1} {\delta A_{Z} \over \delta Z}{d u_{j,m}(Z) \over dZ}
+ u_{j,m}(Z) = 0 .
\end{equation}
If we choose the remaining ordering factor as \( A_{Z}=Z^{2\nu-1} \),
the above equation becomes the Bessel equation
\begin{equation}
{d^{2}u_{j,m}(Z) \over dZ^{2}} - {2\nu-1 \over Z}{du_{j,m}(Z) \over dZ}
+ u_{j,m}(Z) = 0 ,
\label{defmujm}
\end{equation}
and the solution is
\begin{equation}
u^{(\nu)}_{j,m}(Z)=Z^{\nu}[b_{1}H_{\nu}^{(1)}(Z)+b_{2}H_{\nu}^{(2)}(Z)] ,
\label{unuz}
\end{equation}
where \( H_{\nu}(Z) \) is the Hankel function and 
$b_1$ and $b_2$ are constant coefficients. 
In conclusion, the quantum wave function becomes
\begin{equation}
\Psi (Z) = e^{ij\Phi} u^{(\nu)}_{j,m}(Z) \;,
\label{sumsolhank}
\end{equation}
where the argument \(Z \) is expressed using Eqs. (\ref{zett}) and (\ref{fbar}) as
\begin{eqnarray}
Z &=& \int  dr \int ^{\bar{\Lambda} (r)}
       d \bar{\Lambda} \sqrt{ \chi - F_j(\bar{R}) }
       \nonumber \\
  &=& \int dr \biggl(\bar{\Lambda} \sqrt{\chi - F_j(\bar{R})}
     -  {\bar{R}}^{\prime}
\ln \biggl| { \sqrt{\chi} + \sqrt{\chi - F_j(\bar{R})}
        \over \sqrt{\mid F_j(\bar{R}) \mid}}\biggl|
\biggl) \;,
\label{argz}
\end{eqnarray}
where $\chi$ and $F_j$ are given in Eqs. (\ref{chi}) and (\ref{fj}). 

The solution in Eq. (\ref{sumsolhank}) is shown to satisfy 
the original constraint equations (\ref{wd}) automatically, 
because the operators $\hat J$ and $\hat M$ 
are expressed by the linear combination of the original 
constraints (see Eqs. (\ref{angular}) and (\ref{mass})). 
The closure of the algebras among original constraints 
$H$, $H_r$ and $H_{\phi}$ is ensured in a similar way. 


In the final part of this section, 
we comment on other choice of factor ordering for $A_Z$. 
For example, if we choose 
$A_Z=e^{2Z} Z^{-\sigma}(Z-1)^{\sigma}$, 
we obtain the wave function of the hypergeometric function. 
The factor ordering ambiguity appears in our approach, 
because the factor ordering is determined by 
requiring a closed algebra among constraint operators and 
not requiring to define a inner product among state vectors. 

\section{De Broglie-Bohm Interpretation}

We obtained the analytic wave function of the 2+1 dimensional
gravity theory.
The wave function is the functional of the metrics
and its physical and geometrical meaning is not clear.
We adopt the de Broglie-Bohm (dBB) interpretation in order to extract
geometrical information from the wave function.
For this purpose,
we choose only the 2nd kind Hankel function in Eq. (\ref{unuz})
as the wave function which satisfies the Vilenkin boundary condition \cite{Vilenkin82}. 
Then the wave function in the polar coordinates is written as
\begin{eqnarray}
\Psi(Z) =Z^{\nu}H_{\nu}^{(2)}(Z)=\mid \Psi \mid {\rm e}^{i \Theta}(Z) \ .
\label{wafuvil}
\end{eqnarray}
In the dBB interpretation, the momenta are defined by
the gradient of the phase of the wave function as
\begin{eqnarray}
P_{\bar \Lambda} = \frac{\delta \Theta}{\delta \bar \Lambda} \ \ , \ \
P_{\bar R} = \frac{\delta \Theta}{\delta \bar R} \ \ , \ \
P_{\bar C} = \frac{\delta \Theta}{\delta \bar C} \ \ .
\label{dbbmom}
\end{eqnarray}
Inserting the original expressions of the momenta in 
Eqs. (\ref{clmomLambda})- (\ref{clmomC}) and (\ref{canmom}) 
into Eq. (\ref{dbbmom}), 
we obtain the differential equations
for the metrics with respect to the time and space coordinates as
\begin{eqnarray}
\frac{1}{N}(N^r \bar{R}'-\dot{\bar R})&=&\bar{f} \frac{d \Theta}{d Z} \ \ ,
\label{dbb1} \\
\frac{1}{N}((\bar \Lambda N^r)'-\dot{\bar \Lambda})&=&\frac{\bar
\Lambda}{\bar{R}'}
\bar{f}' \frac{d \Theta}{d Z} \ \ ,
\label{dbb2} \\
-\frac{\bar R^3}{N \bar \Lambda} ((N^r {\bar C})'
+ (N^{\phi})'- \dot{\bar C})&=& j \ .
\label{dbb3}
\end{eqnarray}
In the dBB interpretation, the quantum effect is expressed
by the factor
\begin{eqnarray}
(n_{{\rm dBB}})^{-1} := -\frac{d \Theta}{d Z} 
= {2 \over \pi Z |H_{\nu}^{(2)}(Z)|^{2}} \;.
\label{dbbclalim}
\end{eqnarray}
The limit $n_{{\rm dBB}} \rightarrow 1$ is the classical limit, 
because the dBB equations (\ref{dbb1}),(\ref{dbb2}) and (\ref{dbb3}) 
reduce to the Einstein equation in this limit. 
The metrics solving these dBB equations include quantum effects 
through the factor $n_{{\rm dBB}}$. 

In order to integrate the Eqs. (\ref{dbb1})-(\ref{dbb3}),
we fix a part of gauge freedom of
the general coordinate transformation as
\begin{eqnarray}
N^r=N^{\phi}=0 \ .
\end{eqnarray}
Then the metric $\bar C$ is determined from Eq. (\ref{dbb3}) as
\begin{eqnarray}
\dot{\bar C}=\frac{N \bar \Lambda}{\bar R ^3} j \;.
\end{eqnarray}
We can derive the relation free from quantum effects taking
the ratio of Eq. (\ref{dbb2})
over Eq. (\ref{dbb1})
\begin{eqnarray}
\frac{\dot{\bar \Lambda}}{{\bar \Lambda}}=\frac{\dot{\bar R}}{{\bar R^\prime}}
                                          \frac{{\bar f}'}{{\bar f}} \ .
                                          \label{dbb4}
\end{eqnarray}

We show some examples of the solution by
the dBB interpretation in the followings.

\noindent
{\large Example 1: BTZ black hole} \\
If we further restrict the solution by the conditions as
\begin{eqnarray}
{\bar R}'=0 \ \ {\rm and} \ \ \dot{\bar R} = 1 \ ,
\end{eqnarray}
we have the relation $\bar R =t$ .
Then we find
\begin{eqnarray}
ds^2=\frac{1}{F_j}{n_{\rm dBB}}^2dt^2-F_jdr^2+t^2({\bar C}dr+d\phi)^2 \ ,
\end{eqnarray}
from Eqs. (\ref{dbb1}) and (\ref{dbb2})
where $ F_j=1-2m+j^2/(4t^2)-\lambda t^2 \ . $
The classical limit of the metrics under the space-time transformation
$ t\leftrightarrow r , \  \bar C \leftrightarrow N^{\phi} , $
represents the BTZ black \cite{BTZ92} .
\\
{\large Example 2: de Sitter universe} \\
Next we show the example of de Sitter universe.
We set $N=1$ and $m=j=0$ and
we put the ansatz for metrics as
\begin{eqnarray}
\bar{R}=a(t)r \ \ {\rm and}  \ \ \bar \Lambda = b(t)/\sqrt{1-Kr^2} \ ,
\end{eqnarray}
where $K=0$ or $1$ represent the open or closed universe respectively.
From Eq. (\ref{dbb4}), we obtain $a(t)=b(t)$ and
from Eq. (\ref{dbb1}) we obtain
the equation for the scale factor of the universe as
\begin{eqnarray}
\dot a = \sqrt{\lambda a^2 -K} \ { n_{\rm dBB}}^{-1} 
\ .
\label{dota}
\end{eqnarray}
The corresponding classical solutions 
(put $n_{\rm dBB}=1$ in Eq. (\ref{dota})) are open or
closed de Sitter universe
\begin{eqnarray}
   a =\left\{
        \begin{array}{lll}
         \exp{(\sqrt{\lambda}t)}                    & {\rm for} & K=0 \\
         {\lambda}^{-1/2}\cosh{(\sqrt{\lambda}t)}  & {\rm for} & K=1 \ .
        \end{array}
      \right.
\label{clasolsca}
\end{eqnarray}
 The classical closed de Sitter universe
is created with a finite size at $t=0$ and
expands exponentially as $t\gg 0$ .
A special interest is the closed de Sitter universe
because the finite quantum effect can be estimated.

\section{Evaluation of Quantum Effect of de Sitter universe}

In this section, we investigate the quantitative evaluation 
of quantum effects for the metrics of de Sitter universe 
which were studied in the previous section. 
We note that the quantum effect does not appear 
for a open de Sitter universe. 
This is because the argument $Z$ becomes infinity 
as its spatial integration region is infinite 
and therefore the factor $n_{\rm dBB}$ becomes one, 
which indicates the classical limit.  
The finite quantum effect is evaluated only for 
the case of closed de Sitter universe as was explained 
in the previous section: 
\begin{eqnarray}
\bar R=a(t)r \ \ {\rm and } \ \ \bar \Lambda = a(t)/\sqrt{1-r^2} \;,
\end{eqnarray} 
where the scale factor is determined by the dBB equation 
(\ref{dota}) with the relation (\ref{dbbclalim}) as 
\begin{eqnarray}
\dot a = \sqrt{\lambda a^2 -1} \
\frac{2}{\pi Z \mid H_{\nu}^{(2)}(Z) \mid^2 } \ .
\label{dotclosea}
\end{eqnarray} 

\begin{figure}
\epsfxsize=9cm \epsfysize=6cm
\centerline{\epsfbox{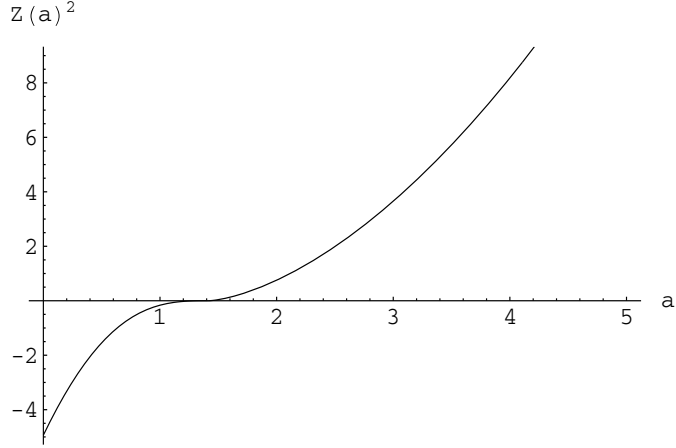}}
\caption{
The argument \(Z^2\) as a function of the scale factor $a$ 
given by Eq. (\ref{arguez})
with the cosmological constant \( \lambda={1 / 2} \)} 
\label{fig:zzaall}
\end{figure}

To evaluate this equation we should first 
integrate with respect to $r$ and obtain 
the expression of the argument $Z$ as a function of the scale 
factor $a(t)$. 
It is expressed in the classical region $(\sqrt{\lambda}a \geq 1)$
and the quantum tunneling region $(0\leq \sqrt{\lambda}a \leq 1)$ 
as 
\begin{eqnarray}
Z&=&
\left\{
\begin{array}{c}
\displaystyle
a \sqrt{\lambda a^{2}-1} 
- {1 \over \sqrt{\lambda}}
\tanh^{-1} {\sqrt{\lambda a^{2} -1} \over \sqrt{\lambda} a}
\ \ \  ({\rm for}  \sqrt{\lambda} a \geq 1) \;,
\\ 
\\
\displaystyle
 i \biggl( a \sqrt{1-\lambda a^{2}}
-{1 \over \sqrt{\lambda}}
\tan^{-1} {\sqrt{1-\lambda a^{2}} \over \sqrt{\lambda} a} \biggl)
\ \  ({\rm for} \; 0 \leq \sqrt{\lambda} a < 1) \;,
\label{imagregzrep}
\end{array}
\right.
\label{arguez}
\end{eqnarray}
The criterion of analytic continuation 
from the classical region 
to the quantum tunneling region is 
that the value of \(Z^{2}\) 
is connected smoothly at boundary point 
\(\sqrt{\lambda}a =1 \), 
because $Z$ takes the real value for the classical region 
and the imaginary value for the quantum region. 
The value of the argument \(Z^2 \) is shown in 
Fig. \ref{fig:zzaall}. 


\begin{figure}
\epsfxsize=9cm \epsfysize=6cm
\centerline{\epsfbox{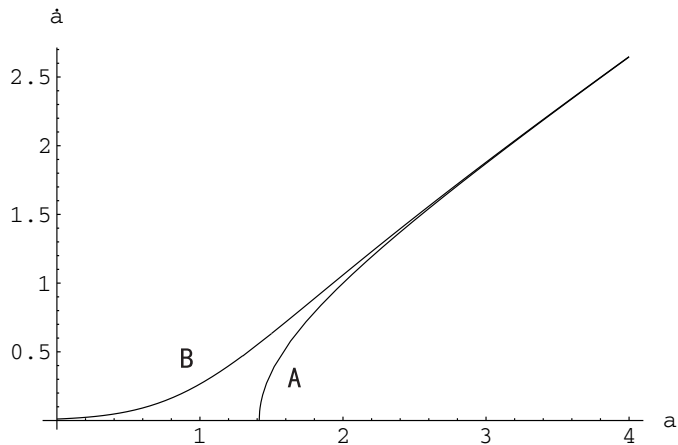}}
\caption{
The rate of change of the scale factor $\dot a(t)$
as a function $a(t)$. 
(A) The classical case: 
\(\dot{a}= \sqrt{\lambda a^2 - 1}\) and 
(B) The quantum case: \(\dot{a}\) in Eq. (\ref{dotclosea}) 
with the cosmological constant \(\lambda={1 / 2}\). 
\label{fig:adotclaqu}}
\end{figure}

\begin{figure}
\epsfxsize=8cm \epsfysize=5cm
\centerline{\epsfbox{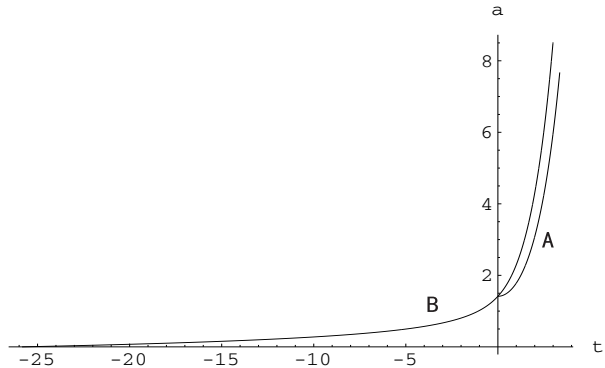}}
\caption{
The scale factor a(t) for the closed de Sitter universe. 
(A) The classical scale factor 
$a(t)={\lambda}^{-1/2}\cosh{(\sqrt{\lambda}t)}$ , 
(B) The numerical solution of Eq. (\ref{defeqat}), 
which includes quantum effects. 
The cosmological constant is set to \( \lambda={1 / 2}\) 
and the time is normalized as $a=1/\sqrt{\lambda}$  at $t=0$. }
\label{fig:allgl33}
\end{figure}

Now we turn to the evaluation of the scale factor 
in Eq. (\ref{dotclosea}). 
We note that at the boundary $\sqrt \lambda a =1$ 
the rate of change of the scale factor $\dot a(t)$ 
is zero for the classical limit ($n_{\rm dBB}=1)$  
and for $| \nu  | > 1/3$ and infinity for $| \nu| < 1/3$, 
where $\nu$ is  the index of the Hankel function. 
The interesting tunneling effect is obtained for 
the case of index of the Hankel function $\nu = 1/3$, 
where the rate of change of the scale factor takes a finite value 
at the boundary: 
\begin{eqnarray}
\lim_{\sqrt {\lambda} a \rightarrow 1} \dot a = 
\lim_{\sqrt {\lambda} a \rightarrow 1}
\sqrt{\lambda a^2 -1} \
\frac{2}{\pi Z \mid H_{1/3}^{(2)}(Z) \mid^2 } 
=\frac{\lambda^{1/6}3^{4/3}\Gamma(2/3)^2}{4 \pi } \ .
\end{eqnarray}
The case for $\nu = - 1/3$ gives the same result because of the 
identity: $H_{-\nu}^{(2)}(Z)=\exp{(-i \pi \nu)}H_{\nu}^{(2)}(Z)$. 
Then the scale factor  
both in the classical region and the quantum region 
is determined by 
\begin{eqnarray}
\dot a(t)=
\left\{
\begin{array}{c}
\displaystyle
\frac{2 \sqrt{\lambda a^2 -1}}{\pi Z | H^{(2)}_{1/3}(Z)|^2}
\ \ \ \ ({\rm for} \; \sqrt{\lambda} a \geq 1) \;,
\\
\\
\displaystyle
\ \frac{2i \sqrt{1-\lambda a^2}}{\pi Z | H^{(2)}_{1/3}(Z)|^2}
\ \ \ \ ({\rm for} \; 0 \leq \sqrt{\lambda} a < 1 ) \;,
\end{array}
\right.
\label{defeqat}
\end{eqnarray}
where the argument $Z$ is given in Eq. (\ref{arguez}).  
From this equation 
the rate of change of the scale factor $\dot a(t) $ 
as a function $a(t)$ is obtained and 
shown in Fig. \ref{fig:adotclaqu}
with \( \lambda = {1 \over 2} \). 
The scale factor as a function of t is obtained 
by integrating Eq.(\ref{defeqat}) numerically 
and the result is shown in Fig. \ref{fig:allgl33}. 

Here we can see from Figs. \ref{fig:adotclaqu} and \ref{fig:allgl33} 
that the quantum scale factor (denoted by a letter B) approaches  
to the classical scale factor (denoted by a letter A) asymptotically 
( $a(t) \rightarrow  \infty $ ), which was required 
by the Vilenkin's boundary condition. 
The quantum scale factor extend 
in the classically forbidden region ($t \leq 0$) 
as a real time quantum tunneling effect. 
This is one of characteristic features of 
the de Broglie-Bohm approach and are 
compared with other approaches: 
the WKB approach \cite{Vilenkin82}
or the path integral approach \cite{HH83}.
A similar analysis was done by 
Horiguchi \cite{Horiguchi94} in 3+1 dimension, 
in which metrics were treated as function of only time $t$ 
while we have treated more generally spherically symmetric metrics 
as function of both time and radial coordinates. 


\section{Summary}

We have studied the 2+1 dimensional 
spherically symmetric gravity theory
and obtained the following results. \\
(1) Quantum theory is defined through local conserved quantities $\hat J(r)$
and
$\hat M(r)$.  \\
(2) The de Broglie-Bohm interpretation is applied for the analytic wave
function of universe. \\
(3) The differential equations of the dBB interpretation are solved and
black hole metrics as well as expanding universe metrics are obtained 
after fixing the coordinate conditions. 
\\
(4) Especially we have evaluated the quantum effects 
of the closed de Sitter universe as the real time tunneling 
in the classically forbidden region occurs 
through the de Broglie-Bohm interpretation.
It is interesting to compare with other approaches
that the universe creates form nothing
through imaginary time tunneling using WKB approximation by Vilenkin
\cite{Vilenkin82}
and the path integral method by Hartle and Hawking \cite{HH83}. 


There are some other interesting topics characteristic in 
2+1 dimensional gravity. 
Localized sources can affect global geometrical structure 
which is expressed as topological conical singularity \cite{DJT84}. 
Other interesting topological solutions are torus solutions 
considered by Hosoya and Nakano \cite{HN90}. 
It may be interesting to make the de Broglie-Bohm interpretation 
for the conical singurarity in the closed universe 
and torus solutions in 2+1 dimensional gravity
because the evaluation of the quantum effects are expected 
to be finite and 
the relation between the quantum solutions and 
the classical solutions becomes clear. 
These are our future problems. 

\section{Acknowledgements}
We would like to thank T.Takahashi for useful discussions and 
numerical calculations. 

\end{document}